\documentclass[twocolumn,letter]{jpsj2}
\def\runauthor{D. Okuyama \textit{et al.}}

\title{
Quadrupolar Frustration in Shastry-Sutherland Lattice of DyB$_4$ Studied by Resonant X-ray Scattering
}

\author{
Daisuke \textsc{Okuyama}\thanks{E-mail address: oku@iiyo.phys.tohoku.ac.jp}, 
Takeshi \textsc{Matsumura}, Hironori \textsc{Nakao} and Youichi \textsc{Murakami} 
}

\inst{
Department of Physics, Graduate School of Science, Tohoku University, Sendai 980-8578\\
}

\recdate{\today}

\abst{
We have observed geometrical frustration of quadrupolar and magnetic moments
in dysprosium tetraboride, DyB$_4$, where the rare-earth sites form a Shastry-Sutherland lattice. 
Resonant X-ray scattering at the $L_{\text{III}}$ absorption edge of Dy was utilized. 
Analysis of the energy, polarization, temperature, and azimuthal-angle dependences
of the $E1$ resonance of the (1 0 0) forbidden reflection show that the magnetic and quadrupolar
components within the frustrated $c$ plane have a short-range correlation, suggesting that
the moments are fluctuating. In contrast, the basic antiferromagnetic component along the $c$-axis
has a long-range order. 
}

\kword{
geometrical frustration, DyB$_4$, resonant X-ray scattering, quadrupole order, fluctuation
}

\begin{document}
\maketitle

Magnetic materials with geometrically frustrated magnetic interactions exhibit a wide
variety of intriguing phenomena caused by fluctuating spins that cannot order down to
very low temperatures. For example, hexagonal ABX$_3$ (A=alkali metal,
B=transition metal, X=halogen) compounds such as CsCoCl$_3$ with a triangular lattice of
Ising spins exhibit an unusual antiferromagnetic state in which one of its three sublattices is
disordered~\cite{Mekata77}.
Frustrations in kagom\'e and pyrochlore lattices offer a platform where a macroscopic number of spin
configurations are degenerate, giving rise to short-range correlated fluctuating spins
called spin liquid states~\cite{Mirebeau05}.
These exotic phenomena are caused, of course, by the spin degree of freedom of the magnetic ions.
On the other hand, there are only a few examples that suggest frustration of the orbital degree of freedom.
In LiNiO$_2$ and FeSc$_2$S$_4$, for example, their anomalous magnetic properties are interpreted
in association with the orbital frustration~\cite{Reynaud01,Fritsch04}.
However, to the best of our knowledge, there has been no experiment in which the frustrated orbitals
were observed directly by an orbital-sensitive method.
In this letter, we present direct evidence for fluctuating quadrupolar and magnetic moments
in a rare-earth compound,  DyB$_4$, which has a Shastry-Sutherland-type geometrical
frustration~\cite{Shastry81}.

\begin{figure}[t]
\begin{center}
\includegraphics*[width=8cm]{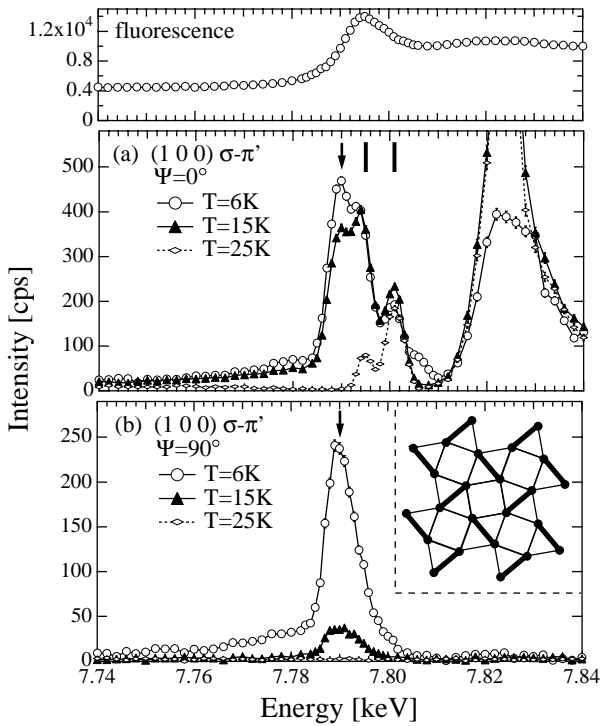}
\caption{
Incident energy dependences of the peak-top intensity of the (1 0 0) reflection for $\sigma$-$\pi'$ at 
(a) $\Psi=0^{\circ}$ and (b) $\Psi=90^{\circ}$. The peak around 7.824 keV for $\Psi=0^{\circ}$ is 
due to the multiple scattering. The top figure shows the fluorescence spectrum.
The inset shows the Dy lattice; the nearest neighbor bonds are shown by the thick lines.
}
\label{fig1}
\end{center}
\end{figure}

Rare-earth tetraborides, RB$_4$, have been attracting growing interest as a system in which the rare-earth network involves a geometrical frustration.
The R ions are at the $4g$ sites of the tetragonal space group of $P4/mbm$:
$\mib{R}_{1}=(x, x+\frac{1}{2}, 0)$, $\mib{R}_{2}=(\frac{1}{2}-x, x, 0)$,
$\mib{R}_{3}=(-x, \frac{1}{2}-x, 0)$, and $\mib{R}_{4}=(\frac{1}{2}+x, -x, 0)$, with $x=0.3175$ for DyB$_4$%
~\cite{Watanuki05a}.
This is equivalent to the Shastry-Sutherland lattice as illustrated in the inset of Fig.~\ref{fig1}(b)%
~\cite{Shastry81}.
It is noted that the nearest-neighbor and next nearest-neighbor distances differ only by 0.33\% 
and that the Dy$^{3+}$ ion has a huge magnetic moment of 10$\mu_{\text{B}}$. 
Therefore, the Shastry-Sutherland lattice of Dy can be regarded as a combination of triangles and squares 
of classical moments. In addition, an orbital degree of freedom is active in DyB$_4$ as described next. 
Consequently, the way of removing frustration in DyB$_4$ is expected to be different from the orthogonal 
dimer system of quantum spins such as SrCu$_2$(BO$_3$)$_2$, 
where the frustration can be removed by making spin-singlet dimers~\cite{Kageyama99,Miyahara99}. 

The physical properties of DyB$_4$ have been studied intensively by
Watanuki \textit{et al}~\cite{Watanuki05a,Watanuki05b}.
Two phase transitions take place, at $T_{\text{N1}}=20.3$ K and at $T_{\text{N2}}=12.7$ K. 
Below $T_{\text{N1}}$, it is established by neutron powder diffraction that an antiferromagnetic order takes place 
with the magnetic moments aligned along the $c$ axis; those on Dy(1) at $\mib{R}_{1}$ and Dy(2) at
$\mib{R}_{2}$ are directed upwards and those on Dy(3) at $\mib{R}_{3}$ and Dy(4) at $\mib{R}_{4}$ downwards.
As a result, forbidden (odd 0 0) reflections arise, but unit cells do not change. 
What is intriguing is that the $C_{44}$ mode of the elastic constant keeps softening even below  $T_{\text{N1}}$. 
The softening, which follows the normal Curie law in phase I ($T\ge T_{\text{N1}}$) that indicates a quadrupolar
degeneracy, is even more enhanced in phase II ($T_{\text{N2}}\le T\le T_{\text{N1}}$).
Moreover, in phase II, strong ultrasonic attenuation occurs, suggesting fluctuation of the quadrupolar moments.
Specific heat measurement shows that the entropy of $R\ln 2$ and $R\ln 4$ is released at $T_{\text{N2}}$ and
at $T_{\text{N1}}$, respectively, with increasing temperature. 
Therefore, the ground state is a pseudo-quartet with quadrupolar degeneracy,
which is somehow not lifted even in phase II. 
In phase III ($T\le T_{\text{N2}}$), the ultrasonic attenuation and softening stop; neutron powder diffraction suggests
that a $c$-plane magnetic component arises in phase III~\cite{MyExperiment05_1}. 
These anomalous behaviors strongly suggest that phase II  is not a normal antiferromagnetic state and 
that there remains fluctuating quadrupolar and magnetic moments,
as first pointed out by Watanuki \textit{et al}~\cite{Watanuki05b}.

In order to investigate phases II and III of DyB$_4$,
we have utilized resonant X-ray scattering (RXS), which has been developed 
as a powerful tool to probe both quadrupolar and magnetic moments~\cite{Matsumura05}. 
Since RXS measures the average structure of the relevant moments in a very short time scale determined by the 
lifetime of the intermediate state ($\sim 10^{-15}$ sec), 
it would be possible to observe a snapshot of the correlated moments
if the fluctuation rate is slower than the time scale of the observation. 
In the present experiment, we have detected a broadened forbidden reflection, indicating
a short-ranged quadrupolar and magnetic order.

A single crystal of DyB$_4$ was grown by the floating zone method using a high-frequency furnace. 
The sample quality was checked in terms of magnetic susceptibility and electrical resistivity, which showed
good agreements with the results in ref.~\citen{Watanuki05a,Watanuki05b}.
Synchrotron X-ray scattering experiments were performed using a four-circle diffractometer 
at beamline BL-16A2 of the Photon Factory in KEK. 
The incident X-ray was tuned near the $L_{\text{III}}$ absorption edge of Dy ($\sim 7.79$ keV),
where the resonance of $2p\leftrightarrow5d$ dipole transition occurs. 
We investigated the ($h$ 0 0) reflections using a sample with the (100) surface. 
Azimuthal-angle scans and polarization analysis using a PG analyzer-crystal were also carried out; 
the azimuthal angle $\Psi$ is defined to be zero when the $c$-axis lies in the scattering plane. 
The mosaic width of the sample was about $0.03^{\circ}$ full width at half-maximum (FWHM) in phase I,
indicating the high quality of the crystal.

Figure \ref{fig1} shows the incident energy dependences of the (1 0 0) forbidden reflection for the 
$\sigma$-$\pi'$ channel at 6 K (phase III), 15 K (phase II), and 25 K (phase I) at two 
azimuthal angles of $\Psi=0^{\circ}$ and $90^{\circ}$~\cite{MyExperiment05_2}. 
The resonant enhancement is clearly observed around the absorption edge. 
We notice three resonant features at $\Psi$=0$^{\circ}$,
which are indicated by an arrow at 7.79 keV and two bars at 7.795 keV and 7.801 keV.
The resonance at 7.79 keV, the inflection point of the fluorescence spectrum, disappears above $T_{\text{N1}}$; 
this can be ascribed to the resonant exchange magnetic scattering originating
from the $4f$ magnetic moment~\cite{Hannon88}.
The other two resonances at 7.795 keV and 7.801 keV, which exist even above $T_{\text{N1}}$, 
can be ascribed to the anisotropic tensor susceptibility (ATS) scattering that reflects the local
crystal-field anisotropy of the surrounding boron atoms.
We call this boron-ATS scattering hereafter.
The double peak energy-spectrum and the $\cos^2 \Psi$ dependence in intensity are the 
same as those in GdB$_4$~\cite{Song03, Lovesey04}. 
At $\Psi$=90$^{\circ}$, the boron-ATS scattering vanishes and only the resonance at 7.79 keV is observed 
as indicated by the arrow. 
We also observed nonresonant reflection below the edge, which we ascribe to magnetic scattering;
the intensity is proportional to the resonance at 7.79 keV. 

\begin{figure}[t]
\begin{center}
\includegraphics*[width=8cm]{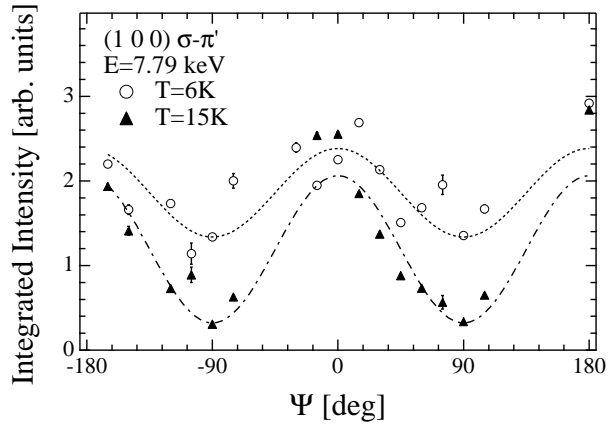}
\caption{
Azimuthal-angle dependence of the integrated intensity of the (1 0 0) resonant reflection 
at $E=7.79$ keV for $\sigma$-$\pi'$. The lines are fits with $\cos^2 \Psi$ and $\sin^2 \Psi$. 
}
\label{fig2}
\end{center}
\end{figure}

We note that the most mysterious point is the existence of the intensity at $\Psi$=90$^{\circ}$ in phase II. 
According to the simple magnetic structure deduced from neutron diffraction, where the antiferromagnetic 
moments 
align completely along the $c$ axis~\cite{Watanuki05a}, the structure factor of the resonant magnetic scattering becomes proportional 
to $\cos \Psi$.
However, this is inconsistent with the data of Fig.~\ref{fig1}(b) for 15 K.
The azimuthal-angle dependence of the resonance intensity shown in Fig.~\ref{fig2} also demonstrates that the 
intensity 
at  $\Psi$=90$^{\circ}$ does not vanish in phase II. 
In phase III, a complex magnetic structure is proposed by neutron scattering, 
where the $c$-plane component of the magnetic moments appears~\cite{Watanuki05a}.
It is therefore expected that the $c$-plane component gives the intesity at $\Psi$=90$^{\circ}$. 

\begin{figure}[t]
\begin{center}
\includegraphics*[width=8cm]{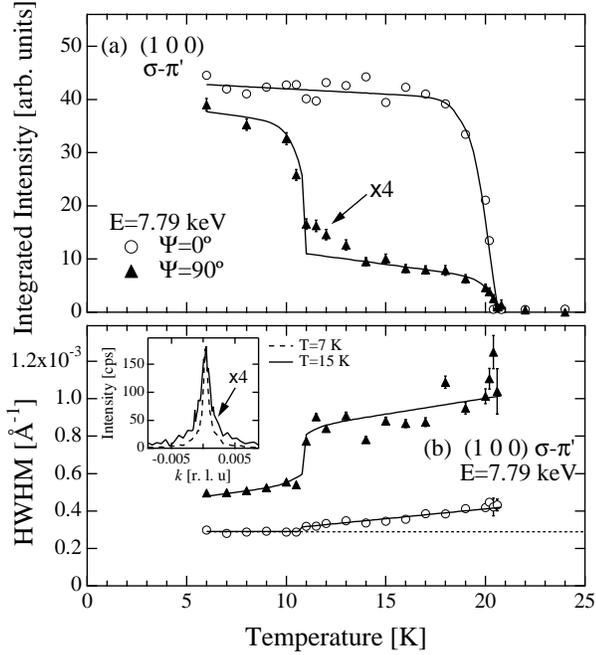}
\caption{
Temperature dependences of (a) the integrated intensity and (b) the HWHM of the rocking scan ($\omega$)
of the (1 0 0) resonant reflection at $\Psi=0^{\circ}$ and $90^{\circ}$ ($E=7.79$ keV, $\sigma$-$\pi'$).
The solid lines are guides for the eye. The dotted line indicates the resolution limit estimated from the fundamental
reflections.
The inset shows the peak profile at 7 K and 15 K for $\Psi=90^{\circ}$.
}
\label{fig3}
\end{center}
\end{figure}

Figure \ref{fig3} shows the temperature dependences of the integrated intensity and half width
at half maximum (HWHM) of the (1 0 0) resonant reflection at 7.79 keV.
There is an obvious difference between the behavior for $\Psi=0^{\circ}$ and $90^{\circ}$.
At $\Psi=0^{\circ}$ the intensity rises up steeply below $T_{\text{N1}}$ and saturates below about 17 K,
whereas at $\Psi=90^{\circ}$ it increases gradually below $T_{\text{N1}}$ and exhibits a sudden
increase below $T_{\text{N2}}$.
Moreover, the HWHM at $\Psi=90^{\circ}$ is obviously broad and exhibits a sudden drop below $T_{\text{N2}}$, 
whereas at $\Psi=0^{\circ}$ it is almost resolution limited and exhibits a slight broadening in phase II. 
It is noted that the rocking ($\omega$) scans in Fig.~\ref{fig3} correspond to the $l$-scan at $\Psi=0^{\circ}$
and $k$-scan at $\Psi=90^{\circ}$ in the $hkl$ reciprocal space.
On the other hand, the HWHM for the $l$-scan at $\Psi=90^{\circ}$ ($\sim$0.01 \AA$^{-1}$ at 16 K and $\sim$0.007 \AA$^{-1}$ at 6 K) is also obviously broader than the resolution ($\sim$0.005 \AA$^{-1}$),
whereas that for the $k$-scan at $\Psi=0^{\circ}$ is resolution limited, although these resolutions are
much worse because the scans correspond to the $\chi$-scan.
These results suggest that a different order parameter exists at $\Psi$=90$^{\circ}$,
which has a short correlation length in phase II.

We briefly refer to the structural phase transition in phase III that has been studied using an X-ray from a Mo target (17.48 keV). 
The single peak of the (0 0 6) reflection at 16 K splits into four peaks along the $h$ and $k$ directions at 9 K;
the (10 0 0) peak becomes broad along the $l$ direction. These results, together with the survey of
some other Bragg peaks, show that the structure changes from tetragonal to monoclinic.
The angle $\angle ac$, or $\angle bc$, is estimated to be 89.84$^{\circ}$ at 9 K.
It should be remarked that the antiferromagnetic domains with the (1 0 0) and (0 1 0) propagation vectors
result in the monoclinic distortions within the $bc$ plane ($\angle bc \neq 90^{\circ}$) and 
$ac$ plane ($\angle ac \neq 90^{\circ}$), respectively;
this is inferred from the fact that the (1 0 2) resonant magnetic Bragg peak splits only
along the $k$ direction in phase III.

Let us discuss the origin of the nonzero intensity of the broad (1 0 0) resonant reflection at $\Psi=90^{\circ}$ in phase II. 
We analyze the RXS result by taking into account the $c$-plane components of both magnetic (rank 1)
and quadrupolar (rank 2) moments. 
For this purpose, we use a theory developed by Lovesey \textit{et al.} which directly connects the atomic tensors to the
scattering amplitude of RXS~\cite{Matsumura05,Lovesey01}.

\begin{figure}[t]
\begin{center}
\includegraphics*[width=8cm]{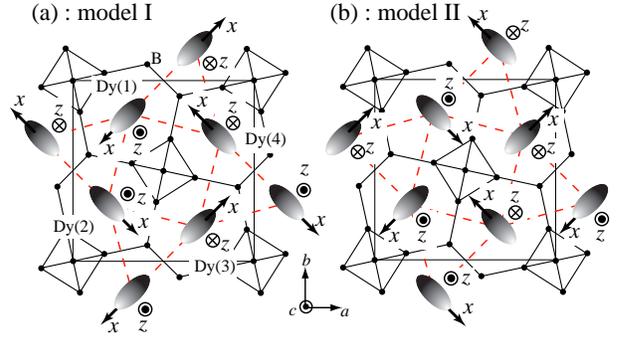}
\caption{
Models of possible magnetic and quadrupolar structures.
The local $x$- and $z$-axes correspond to the $c$-plane and $c$-axis magnetic components 
of each Dy ion, respectively.
Ellipses with light and dark shades represent $\langle O_{zx}\rangle$-type charge distributions 
that are extended above and below the plane of the paper, respectively.
}
\label{fig4}
\end{center}
\end{figure}

Here, we examine two possible models of the in-plane magnetic and quadrupolar structures that
are consistent with the experimental results.
They are illustrated in Fig.~\ref{fig4}. 
We take the directions of the $c$-plane and $c$-axis components as the local $x$- and $z$-axes, respectively. 
The atomic tensors expressed in the local $xyz$ coordinates are assumed to be the same for 
all the four Dy ions; they can be transformed to each other by appropriate Euler rotations.
Quadrupolar moments represented by the ellipses can coexist because of the strong spin-orbit interaction.

The unit cell structure factor of the ($h$=odd 0 0) reflection for the two models in Fig.~\ref{fig4} is written as
\begin{eqnarray}
F^{(E1)}_{\sigma\text{-}\pi'} \!\!\!&=&\!\!\!
2k_{1}b\cos{\theta}(\sqrt{2}\langle J_{z}\rangle\cos{\Psi}-\langle J_{x}\rangle
\sin{\Psi}) \nonumber\\
&&\!\!\!\!\!\!\!\!\mp2k_{2}a\cos{\theta}(\sqrt{2}\langle O_{22}\rangle\cos{\Psi}+\langle O_{zx}\rangle
\sin{\Psi})
\label{eq:1}
\end{eqnarray}
for $E1$ resonance and the $\sigma$-$\pi'$ channel, 
where the minus sign in the second line corresponds to model I and the plus sign to model II; 
$a=\cos (2\pi h x)$ and $b=\sin (2\pi h x)$; $\theta$ is the Bragg angle; 
$k_1$ and $k_2$ are constant real numbers. 
Note that the tensors in eq.~(\ref{eq:1}) are of the $5d$ state; 
$\langle J_{\alpha} \rangle$ can be induced through magnetic exchange interaction 
with $\langle J_{\alpha} \rangle_{4f}$; $\langle O_{\gamma} \rangle$ through 
$d$-$f$ Coulomb interaction with $\langle O_{\gamma} \rangle_{4f}$ or through 
mixing with the $2p$ orbitals of boron atoms (boron-ATS mechanism)\cite{Matsumura05}.
The boron-ATS mechanism obviously gives rise to $\langle O_{22} \rangle$, and this term explains
the azimuthal-angle dependences of the resonances at 7.795 keV and 7.801 keV in phase I.
In phase II, the $c$-axis component of the $4f$ magnetic moment induces $\langle J_z \rangle$. 
The $\langle J_z \rangle$ term explains the $\cos^2 \Psi$ component in the azimuthal-angle dependence 
of the resonance at 7.79 keV as shown in Fig.~\ref{fig2}. 

The $\langle J_x \rangle$ and $\langle O_{zx} \rangle$ terms, with the $\sin{\Psi}$ dependence,
can explain the nonzero intensity at $\Psi$=90$^{\circ}$. 
However, because the $\cos\Psi$ and $\sin\Psi$ terms interfere coherently in eq.~(\ref{eq:1}),
the structure factor can be transformed into the form $A\cos(\Psi+\alpha)$, where $A$ and $\alpha$ are
constants.
This means that the intensity must vanish somewhere in the azimuthal scan, which is the case in GdB$_4$~\cite{Song03}.
However, the experimental data in Fig.~\ref{fig2} do not support this simple scenario. 
The data suggest that the $\cos\Psi$ and $\sin\Psi$ terms contribute independently 
rather than interfere coherently. This is demonstrated by the lines in Fig.~\ref{fig2}. 
It seems that the resonance at $\Psi=90^{\circ}$ represents the in-plane components
$\langle J_x \rangle$ and $\langle O_{zx} \rangle$, and the resonance at 
$\Psi=0^{\circ}$ represents the $\langle J_z \rangle$ and $\langle O_{22} \rangle$ moments, respectively. 

A possible reason for this behavior 
is that the correlation length the of $\langle J_x \rangle$
and $\langle O_{zx} \rangle$ components is much shorter than that of $\langle J_z \rangle$
and $\langle O_{22} \rangle$. 
By fitting the $k$-scan profile at $\Psi=90^{\circ}$ with a Lorentzian that is convoluted with the resolution function,
the correlation length of $\langle J_x \rangle$ and $\langle O_{zx} \rangle$ along the $b$-axis
is estimated to be $(1.4\pm 0.1)\times 10^3$ \AA\ at 15 K in phase II and
$(4.2\pm0.2)\times 10^3$ \AA\ at 7 K in phase III.
The correlation length along the $c$-axis is also short, less than the resolution
of $1.5\times 10^3$ \AA\, although precise estimation was difficult. 
On the other hand, the correlation of $\langle J_z \rangle$ and $\langle O_{22} \rangle$ along the $c$-axis
estimated from the $l$-scan profile at $\Psi=0^{\circ}$ is $(9.2\pm0.5)\times 10^3$ \AA\ at 15 K
and over $2\times 10^4$ \AA, the resolution limit, at 7 K. The correlation along the $b$-axis
is also expected to be longer than the resolution limit of $1.5\times 10^3$ \AA.
Then, the volume fraction from which the scattering occurs coherently is much smaller for 
$\langle J_x \rangle$ and $\langle O_{zx} \rangle$ than for $\langle J_z \rangle$ and $\langle O_{22} \rangle$.
Therefore, the $\cos\Psi$ and $\sin\Psi$ terms in eq.~(\ref{eq:1}) almost do not interfere and
they give reflections independently. 

Thus, we consider that the resonance at $\Psi=90^{\circ}$ reflects $\langle J_x \rangle$ or
$\langle O_{zx} \rangle$, which is induced by the magnetic or quadrupolar moments
of the $4f$ electrons, respectively.
The peak profiles are clearly broadened in phase II, indicating that the correlation length is short.
Furthermore, these in-plane components are considered to be fluctuating.
This is supported by neutron powder diffraction in phase II,
which shows that $\langle gJ_z \rangle_{4f}$ is only 6.9$\mu_{\text{B}}$ while the full moment is 10$\mu_{\text{B}}$~\cite{Watanuki05a}.
The strong ultrasonic attenuation in the $C_{44}$ mode also supports the fluctuation of $\langle O_{zx} \rangle$~\cite{Watanuki05b}.
In phase III, neutron diffraction shows that an in-plane component of 5.3$\mu_{\text{B}}$ appears,
resulting in a total moment of 8.8$\mu_{\text{B}}$; the ultrasonic attenuation and softening also stop.
RXS at $\Psi$=90$^{\circ}$ shows that the correlation length becomes longer, although it is not 
as long as the resolution limit. 
These results show that the fluctuation stops in phase III and a static short-range order of
$\langle J_x \rangle$ and $\langle O_{zx} \rangle$ remains. 

Which of $\langle J_x \rangle$ and $\langle O_{zx} \rangle$ is dominant
in this reflection?
Unfortunately, it is not distinguishable from the present experimental results and from eq.~(\ref{eq:1})
because the two factors are completely in phase.
In the present case of DyB$_4$,
because $\langle J_z \rangle$ is finite,
$\langle O_{zx} \rangle$ will be induced if $\langle J_x \rangle$ arises because of the strong spin-orbit coupling,
and $\langle J_x \rangle$ will be induced if $\langle O_{zx} \rangle$ arises as well.
Ultrasonic attenuation and softening of the $C_{44}$ mode seem to support the dominance of
$\langle O_{zx} \rangle$.
It would be possible to distinguish if we could investigate by RXS 
the relative change in the signal for $h=$1, 3, and 5 of the ($h$ 0 0) reflection
through changes in the factors $a$ and $b$, which have different $h$ dependences. 

With respect to the origin of the short-range correlation and fluctuation of the in-plane moment, 
one plausible scenario is that the $\langle O_{zx} \rangle$ moments in the triangular connection of the 
Dy lattice experience frustration. 
If we assume the charge distribution of the $4f$ electrons prefer to avoid each other, 
the arrangement of $\langle O_{zx} \rangle$ in models I and II experiences frustration, which could lead to an orbital liquid state. 

We consider that the static order of the in-plane  moments takes place when the frustration is removed
by the lattice distortion.
The arrangement of the $\langle O_{zx} \rangle$-type quadrupolar moment in models I and II will favor
a uniform lattice distortion through a cooperative Jahn-Teller effect into a monoclinic structure where the
$bc$ plane is distorted ($\angle bc \neq 90^{\circ}$).
This is consistent with the experimental result in phase III.
A possible space group of the monoclinic phase allowed by Landau theory is $P2_1/c$,
where all the atomic sites are described by the $4e$ site~\cite{Stokes88}.
It is intriguing that the peak width at $\Psi=90^{\circ}$ is still broader than the resolution limit,
which means that the correlation length does not diverge.

In summary, we have investigated the phase transitions occuring in the Shastry-Sutherland lattice of DyB$_4$
by resonant X-ray scattering, and detected the short-range correlated in-plane moments
$\langle J_{x} \rangle$ and $\langle O_{zx} \rangle$.
They are expected to be fluctuating in the high-temperature antiferromagnetic phase because of frustration
and they change into possessing a static order with short-range correlation in the low-temperature phase with monoclinic distortion.
The present results indicate that the doublet ground state remains
well below $T_{\text{N1}}$ despite the magnetic ordering, at which the $c$-axis
component is fixed and the pseudo-quartet should be split into four
singlets by the Zeeman effect.  Why does the doublet exist in Phase
II? What kind of quadrupolar degeneracy is realized?  Such fundamental
issues need to be solved in future studies.

The authors are indebted to R. Watanuki for many fruitful discussions.
We also thank S. Kunii, K. Horiuchi, M. Onodera, H. Shida for assistance in single-crystal growth,
Y. Wakabayashi for experimental support, and K. Iwasa and K. Suzuki for profitable discussions.
This study was performed with the approval of the Photon Factory Program Advisory
Committee (No. 2004G235), and was supported by the 21st century center of excellence program, 
and by a Grant-in-Aid for Scientific Research from the Japanese Society for the Promotion of Science.


\begin{thebibliography}{99}
\bibitem{MyExperiment05_1} We call the magnetic components parallel and perpendicular to
the $c$-axis as the $c$-axis component and $c$-plane component, respectively.
\bibitem{MyExperiment05_2} No reflection was observed for $\sigma$-$\sigma'$ at any azimuthal angle.
\end{thebibliography}
\end{document}